\documentstyle[11pt]{article}
\newcommand{\be}{\begin{equation}}
\newcommand{\ee}{\end{equation}}
\newcommand{\bea}{\begin{array}}
\newcommand{\eea}{\end{array}}

\title{WDVV and DZM}
\author{Robert Carroll\\
Mathematics Dept.\\
University of Illinois,
Urbana, IL 61801\\email:  rcarroll@symcom.math.uiuc.edu}

\date{March, 1997}

\begin{document}
\bibliographystyle{plain}
\maketitle

\begin{abstract} 
We show how the 
Witten-Dijkgraff-Verlinde-Verlinde (WDVV) equations and the 
Darboux-Zakharov-Manakov (DZM) system can be characterized via a
background family of functions.
\end{abstract}


\section{INTRODUCTION}
\renewcommand{\theequation}{1.\arabic{equation}}\setcounter{equation}{0}

The background
literature for WDVV in terms of topological field theory (TFT)
goes back to \cite{da,wa} for example and an
extensive development appears in \cite{db}, connecting the equations
to Frobenius manifolds and Egorov geometry.  A recent paper \cite{ka}
develops this point of view on Riemann surfaces and other recent work
in \cite{bf,ma} connects matters to $N=2$ susy Yang-Mills (YM) or
Seiberg-Witten (SW) theory.  On the other hand the DZM system goes
back to \cite{na,za} for example and more recently there have been 
extensive developments in \cite{bd} (cf. also \cite{ba,bb,bc,bi}).
We will exhibit here some of the connections between
WDVV and DZM in a somewhat different abstract manner which reveals
the purely algebraic character of certain features.  For simplicity
we do not give a survey of background ideas on Egorov geometry from
\cite{db,dc,ka} but will mention some points of contact as we go along
(cf. \cite{cz,kc,md} for an extended treatment of these matters in a broader
context).

\section{DZM}
\renewcommand{\theequation}{2.\arabic{equation}}\setcounter{equation}{0}

To see how the DZM theory can arise independently in perhaps maximum
generality we follow \cite{bd}.
(cf. also \cite{ba,bb,bc,bd,bi,cd,kb,zb}). 
Thus first a background situation here goes back to
\cite{cd,kb} where the Hirota bilinear identity was derived from the D-bar
framework.  This connection involves algebraic techniques from Sato
theory on one side and analytic techniques from D-bar on the other.
Connections based on Hirota as in \cite{cd,kb}, or more generally
in \cite{bd}, form a bridge or marriage between the two types
of technique and touch upon the intrinsic meaning of the whole business.
Now for the background derivation of \cite{cd,kb} we consider 
the (matrix) formula
\be
\bar{\partial}_{\lambda}\psi(x,\lambda,\bar{\lambda})=\int\int_{\Omega}\psi
(x,\lambda',\bar{\lambda}')R_0(\lambda',\bar{\lambda}',\lambda,
\bar{\lambda})\,d\lambda'\wedge d\bar{\lambda}';
\label{BA}
\ee
$$\bar{\partial}_{\lambda}\tilde{\psi}(x',\lambda,\bar{\lambda})=
-\int\int_{\Omega}R_0(\lambda,\bar{\lambda},\lambda',\bar{\lambda}')
\tilde{\psi}(x',\lambda',\bar{\lambda}')\,d\lambda'\wedge d\bar{\lambda}'$$
Multiply by $\tilde{\psi}(x',\lambda,\bar{\lambda})$ on the right in the
first equation and by $\psi(x,\lambda,\bar{\lambda})$ on the left in the
second to obtain
\be
\int\int_{\Omega}\bar{\partial}[\psi(x,\lambda,\bar{\lambda})\tilde
{\psi}(x',\lambda,\bar{\lambda})]\,d\lambda\wedge d\bar{\lambda}=
-\int_{\partial\Omega}\psi(x,\lambda,\bar{\lambda})\tilde{\psi}
(x',\lambda,\bar{\lambda})d\lambda=0
\label{BB}
\ee
which is the Hirota bilinear identity when $\partial\Omega\sim C$ is a small
circle around $\infty$.
\\[3mm]\indent
We go next to \cite{bb,bd,bi,zb} (a complete discussion of
this with details and derivations will appear in \cite{cz}).
Here we go directly to 
\cite{bd} and take
$(\spadesuit\spadesuit)\,\,\psi(\lambda,\mu,g)=g^{-1}(\mu)\chi(\lambda,\mu,g)
g(\lambda)$
with $\eta=(\lambda-\mu)^{-1}$ and
$g\sim g_i=exp(\sum K_ix_i)$ where $K_i(\lambda)$ are
commuting meromorphic matrix functions.  It is assumed now that there
is some region $G\subset {\bf C}$ where $R(\lambda,\mu)=0$ in $G$
with respect to $\lambda$ and $\mu$ (we will take this to mean that $R=0$
whenever $\lambda$ or $\mu$ are in $G$).  Also $G$ contains all zeros
and poles of the $g(\lambda)$ and $G$ contains a neighborhood of $\infty$.
For $K_i(\lambda)$ a polynomial in $\lambda$ this involves only $\{
\infty\}\subset G$ whereas for $g=(\lambda -a)^{-1}$ it requires 
$\{a,\infty\}\subset G$.  Some examples are also used where $g\sim
exp(\sum x_i
\lambda^{-i})$ with $G$ a unit disc.  For now we think of $G$ as some region
containing $\infty$ and $\eta=(\lambda-\mu)^{-1}$ which leads to
($g_1\sim g(\lambda,x)$ and $g_2\sim g(\lambda,x')$)
\be
\bar{\partial}_{\lambda}\chi(\lambda,\mu)=2\pi i\delta(\lambda-\mu)+
\int_{{\bf C}}d^2\nu\,\chi(\nu,\mu)g_1(\nu)R(\nu,\lambda)g_1(\lambda)^{-1};
\label{BN}
\ee
$$\bar{\partial}_{\lambda}\chi^*(\lambda,\mu)=2\pi i\delta(\lambda-\mu)
-\int_{{\bf C}}d^2\nu\, g_2(\lambda)R(\lambda,\nu)g_2(\nu)^{-1}\chi^*
(\nu,\mu)$$
One can also take $R(x,\lambda',\lambda)=g(x,\lambda')R_0(\lambda',\lambda)
g^{-1}(x,\lambda)$ and in (\ref{BN}) we should think of $R$ as an $R_0$ 
term and write (\ref{BN}) as
\be
\bar{\partial}_{\lambda}\left[g_1(\mu)\psi(\lambda,\mu,g_1)g_1^{-1}(\lambda)
\right]=2\pi i\delta(\lambda-\mu)+
\label{BP}
\ee
$$+\int d^2\nu\left[g_1(\mu)\psi(\nu,\mu,g_1)
g_1^{-1}(\nu)\right]g_1(\nu)R_0(\nu,\lambda)g_1^{-1}(\lambda)
$$
Then $\tilde{R}(\nu,\lambda)=g_1(\nu)R_0(\nu,\lambda)g_1^{-1}(\lambda)$ plays
the role of $R$ in \cite{cd} and (\ref{BP})
becomes for $g_1$ analytic
\be
\bar{\partial}_{\lambda}\psi(\lambda,\mu,g_1)=2\pi ig_1^{-1}(\mu)\delta
(\lambda-\mu)g_1(\lambda)+\int d^2\nu\,\psi(\nu,\mu,g_1)R_0(\nu,\lambda)
\label{BQ}
\ee(actually $\tilde{R}\sim\tilde{R}_1$ here). One
can also stipulate an equation
$(\bullet)\,\,\partial_i
\tilde{R}(\lambda,\mu,x)=K_i(\lambda)\tilde{R}-\tilde{R}K_i(\mu)$.
Some calculations now give $(\clubsuit\clubsuit)
\,\,\chi^*(\mu,\lambda,g)=-\chi(\lambda,\mu,g)$
and, generally and
aside from the formulas (\ref{BN}), we know that the functions
$\chi,\,\chi^*$, and the $g_i$ are analytic in ${\bf C}/G$ so by
Cauchy's theorem ($\lambda,\,\mu\in G$)
\be
0=-\int_{\partial G}\chi(\nu,\lambda,g_1)g_1(\nu)g_2^{-1}(\nu)\chi^*(\nu,\mu)
d\nu=\int_{\partial G}\chi(\nu,\lambda)g_1(\nu)g_2^{-1}(\nu)\chi(\mu,\nu)d\nu
\label{BS}
\ee
Note here that from (\ref{BN}) one knows $\chi(\lambda,\mu)\sim
(\lambda-\mu)^{-1}$ for $\lambda\to\mu$ and $\chi^*(\lambda,\mu)\sim
(\lambda-\mu)^{-1}$ as well.  Then from (\ref{BS}) and a residue calculation
one obtains for $g_1=g_2$ another proof of $(\clubsuit\clubsuit)$,
requiring only that the $g_i$ be analytic in ${\bf C}/G$.  
Finally we can write (\ref{BS}) in terms of $\psi$ via
$(\spadesuit\spadesuit)$,
namely
\be
0=\int_{\partial G}\psi(\nu,\lambda,g_1)\psi(\mu,\nu,g_2)d\nu
\label{BT}
\ee
which is a more familiar form of Hirota bilinear identity (but now
generalized considerably).
\\[3mm]\indent
One can derive the DZM equations immediately from
the Hirota bilinear identity (\ref{BS}) as in \cite{bd}. 
Thus write $g(\nu)=exp[K_i(\nu)x_i]$ with $K_i=
A_i(\lambda-\lambda_i)^{-1}$ so that $g_1(\nu)g_2^{-1}(\nu)=exp[A_i
(\nu-\lambda_i)^{-1}(x_i-x_i')]$.  Look at (\ref{BS}) and differentiate
in $x_i$ (with $x_i'$ fixed); then let $x_i'\to x_i$ to obtain
\be
0=\int_{\partial G}\left[\partial_i\chi(\nu,\mu,g_1)]\chi(\lambda,\nu,g_1)+
\chi(\nu,\mu,g_1)\frac{A_i}{\nu-\lambda_i}\chi(\lambda,\nu,g_1)\right]d\nu
\label{GE}
\ee
Computing residues yields ($g\sim g_1$)
\be
-\partial_i\chi(\lambda,\mu,g)+\chi(\lambda_i,\mu,g)A_i\chi(\lambda,\lambda_i,g)
+\frac{A_i}{\mu-\lambda_i}\chi(\lambda,\mu,g)-\chi(\lambda,\mu,g)\frac{A_i}
{\lambda-\lambda_i}=0
\label{GF}
\ee
Using the
relation $(\spadesuit\spadesuit)\,\,\psi(\lambda,\mu,g)=g^{-1}(\mu)
\chi(\lambda,\mu,g)g(\lambda)$ this is immediately seen to be equivalent to
\be
\partial_i\psi(\lambda,\mu,g)=\psi(\lambda_i,\mu,g)A_i\psi(\lambda,\lambda_i,g)
\label{GG}
\ee
To derive the DZM system take for $G$ a set of three identical unit discs 
$D_i$ with
centers at $\lambda=0$.  The functions $K_i(\lambda)$ have the form
$K_i(\lambda)=A_i/\lambda$ for $\lambda\in D_i$ and $K_i(\lambda)=0$
for $\lambda\not\in D_i$ (the $A_i$ are commuting matrices).  Evaluating
(\ref{GG}) for independent variables 
$\lambda,\,\mu\in \{0_i,0_j,0_k\}$ one obtains
\be
\partial_i\psi(\lambda,\mu)=\psi(0_i,\mu)A_i\psi(\lambda,0_i),\,\,
\partial_i\psi(\lambda,0_j)=\psi(0_i,0_j)A_i\psi(\lambda,0_i);
\label{GH}
\ee
$$\partial_i\psi(0_j,\mu)=\psi(0_i,\mu)A_i\psi(0_i,0_j);\,\,
\partial_i\psi(0_j,0_k)=\psi(0_j,0_i)A_i\psi(0_i,0_k)$$
Now one can integrate equations containing $\lambda,\,\mu$ over $\partial G$
with weight functions $\rho(\lambda),\tilde{\rho}(\mu)$ so there results
(no sum over repeated indices)
\be
\partial_i\Phi=\tilde{f}_if_i;\,\,\partial_if_j=\beta_{ji}f_i;\,\,
\partial_i\tilde{f}_j=\tilde{f}_i\beta_{ij};\,\,\partial_i\beta_{jk}
=\beta_{ji}\beta_{ik}
\label{GI}
\ee
where
\be
\Phi=\int\tilde{\rho}(\mu)\psi(\lambda,\mu)\rho(\lambda)d\lambda d\mu;\,\,
\beta_{ij}=(A_j)^{1/2}\psi(0_j,0_i)A_i^{1/2};
\label{GJ}
\ee
$$f_i=A_i^{1/2}\int\psi(\lambda,0_i)\rho(\lambda)d\lambda;\,\,
\tilde{f}_i=\int\tilde{\rho}(\mu)\psi(0_i,\mu)A_i^{1/2}d\mu$$
The system of equations (\ref{GI}) implies that
\be
\partial_i\partial_j\tilde{f}_k=[(\partial_j\tilde{f}_i)\tilde{f}_i^{-1}]
\partial_i\tilde{f}_k+[(\partial_i\tilde{f}_j)\tilde{f}_j^{-1}]\partial_j
\tilde{f}_k;
\label{GK}
\ee
$$\partial_i\partial_j\Phi=[(\partial_j\tilde{f}_i)\tilde{f}_i^{-1}]
\partial_i\Phi+[(\partial_i\tilde{f}_j)\tilde{f}_j^{-1}]\partial_j\Phi$$
\be
\partial_i\partial_jf_k=(\partial_if_k)(f_i^{-1}\partial_jf_i)+
(\partial_jf_k)(f_j^{-1}\partial_if_j);
\label{GL}
\ee
$$\partial_i\partial_j\Phi=\partial_i\Phi f_i^{-1}(\partial_jf_i)+
\partial_j\Phi f_j^{-1}(\partial_if_j)$$
The first system in (\ref{GK}) is the matrix DZM equation with the
first system in (\ref{GL}) as its dual partner.
At this stage the development is purely abstract; no reference to
Egorov geometry or TFT is involved.
In this note we will refer to 
\be
\partial_if_j=\beta_{ji}f_i\,\,(i\not=j);\,\,\partial_i\beta_{jk}=
\beta_{ji}\beta_{ik}\,\,(i\not=j\not=k);\,\,\beta_{ij}=\beta_{ji};
\label{WA}
\ee
as a (reduced) DZM system (see Remark 3.1 for the last condition). 
In addition one will want a condition
\be
\partial\beta_{ij}=\partial f_j=0\,\,\,(\partial=\sum\partial_k)
\label{WAA}
\ee
discussed below (cf. Remark 3.1 in particular).  Note that 
\be
\partial\beta_{jk}=\sum_i\beta_{ji}\beta_{ik}=B^2_{jk}
\label{WAB}
\ee
so (\ref{WAA}) for $\beta_{ij}$ implies that $B^2=0$ where $B=(\beta_{ij})$.
One recalls also that 
\be
\partial_k\beta_{ij}=\beta_{ik}\beta_{kj}\,\,(i\not=j\not=k);\,\, 
\partial\beta_{ij}\equiv
\partial_i\beta_
{ij}+\partial_j\beta_{ji}+\sum_{m\not=i,j}\beta_{im}\beta_{mj}=0\,\,(i\not=j)
\label{WB}
\ee
are referred to as Lam\'e equations.  They correspond to vanishing conditions
$R_{ij,ik}=0$ and $R_{ij,ij}=0$ respectively for the curvature tensor
of the associated Egorov metric (see Section 3).
Compatibility conditions
for the equations (\ref{GL}) give the equations 
for rotation coefficients $\beta_{ij}$
as in (\ref{GI}).  One has the freedom to choose the weight function
$\tilde{\rho}$ keeping the rotation coefficients invariant, and
this is described by the Combescure symmetry transformation 
\be
(\tilde{f}'_i)^{-1}\partial_i\tilde{f}'_j=\tilde{f}_i^{-1}\partial_i
\tilde{f}_j
\label{GM}
\ee
Similarly the dual DZM system admits $(\partial_if_j')(f'_i)^{-1}=
(\partial_if_j)f_i^{-1}$.  The function $\Phi$ is considered as a wave
function for two linear problems (with different potentials) corresponding
to the DZM and dual DZM systems.  A general Combescure transformation
changes solutions for both the original system and its dual (i.e.
both $\rho$ and $\tilde{\rho}$ change).
We note also that,
according to \cite{ta,tb}, the theory of Combescure transformations 
coincides with the theory of integrable diagonal systems of hydrodynamic
type.  
\\[3mm]\indent
It is natural now to ask whether some general WDVV equations 
(see below) arise
directly from DZM as formulated here, without explicit reference
to Egorov geometry etc. 
We emphasize however that the role of Egorov geometry and its
many important connections to integrable systems,
TFT, etc. is fundamental here (cf. Section 3); we are
in fact using the Egorov geometry to isolate
some algebraic features, after which the geometry disappears (cf. here
\cite{ma} for example where the need for a general context is indicated).  
In this direction 
consider a scalar situation where $f_j\sim\psi_j$ and assume $\beta_{ij}
=\beta_{ji}$ in (\ref{GI}) as indicated in (\ref{WA}).  Then
$\partial_if_j=\beta_{ji}f_i$ and 
$\partial_jf_i=\beta_{ij}f_j$ which implies $\partial_if^2_j=\partial_j
f^2_i$.  This corresponds to the existence of a function $G$ such that
$f^2_i=\partial_iG=G_i$.
Then look at (\ref{GL}) where (no sums)
\be
f_k\partial_i\partial_jf_k=\frac{f_kf_i\partial_if_k\partial_jf_i}{f^2_i}+
\frac{f_kf_j\partial_jf_k\partial_if_j}{f_j^2}\Rightarrow
\label{JA}
\ee
$$\Rightarrow f_k\partial_i\partial_jf_k=\frac{G_{ki}G_{ij}}{4G_i}+\frac
{G_{kj}G_{ji}}{4G_j}$$
Also from $2f_k\partial_jf_k=G_{kj}$ one gets $2\partial_if_k\partial_jf_k+
2f_k\partial_i\partial_jf_k=G_{kji}$.  Similarly
$2f_i\partial_jf_i=G_{ij}$ implies $2\partial_kf_i\partial_jf_i+2f_i
\partial_k\partial_jf_i=G_{ijk}$ and $2f_j\partial_if_j=G_{ji}$ implies
$G_{jik}=2\partial_kf_j\partial_if_j +2f_j\partial_k\partial_if_j$.  Hence,
using (\ref{JA}), we get
\be
2f_k^2G_{ijk}=4f_k^2\partial_if_k\partial_jf_k+4f_k^3\partial_i\partial_jf_k=
G_{ik}G_{jk}+f_k^2\left[\frac{G_{ki}G_{ij}}{G_i}+\frac{G_{kj}G_{ji}}{G_j}
\right]
\label{JB}
\ee
This implies
\be
2G_{ijk}=\frac{G_{ik}G_{jk}}{G_k}+\frac{G_{ki}G_{ij}}{G_i}
+\frac{G_{kj}G_{ji}}{G_j}
\label{WC}
\ee
Similarly, $\tilde{f}^2_i=\partial_i\tilde{G}$ and
\be
\partial_i\Phi=\tilde{f}_if_i\Rightarrow f_i\tilde{f}_i\partial_j\partial_i
\Phi=\frac{1}{2}\left(G_i\tilde{G}_{ij}+\tilde{G}_iG_{ij}\right)
\label{JC}
\ee
while from (\ref{GK}) - (\ref{GL}) one has
\be
\partial_j\partial_i\Phi=\left\{
\begin{array}{c}
f_i\partial_j\tilde{f}_i+f_j\partial_i\tilde{f}_j\\
\tilde{f}_i\partial_jf_i+\tilde{f}_j\partial_if_j
\end{array}
\right.
\label{JD}
\ee
Consequently
\be
f_if_j\partial_i\partial_j\Phi=\frac{1}{2}(f_i\tilde{f}_j+f_j\tilde{f}_i)G_{ij};
\label{JE}
\ee
$$\tilde{f}_i\tilde{f}_j\partial_i\partial_j\Phi=\frac{1}{2}(f_i\tilde{f}_j
+f_j\tilde{f}_i)\tilde{G}_{ij}$$
Further 
\be
\beta_{ij}=\beta_{ji}=\frac{\partial_if_j}{f_i}=\frac{\partial_jf_i}{f_j}
\Rightarrow
f_if_j\beta_{ij}=\frac{1}{2}G_{ij};\,\,\tilde{f}_i\tilde{f}_j
\beta_{ij}=\frac{1}{2}\tilde{G}_{ij}
\label{JF}
\ee
Note also from (\ref{GI}) $\partial_i\beta_{jk}
=\beta_{ji}\beta_{ik},\,\,\partial_k\beta_{ij}=\beta_{ik}\beta_{kj},$ and
$\partial_j\beta_{ik}=\beta_{ij}\beta_{jk}$, while from (\ref{JF}) we have
\be
(\partial_kf_i)f_j\beta_{ij}+f_i(\partial_kf_j)\beta_{ij}+f_if_j\partial_k
\beta_{ij}=\frac{1}{2}G_{ijk}
\label{JG}
\ee
This leads to
\be
\frac{1}{2}G_{ijk}=\left\{
\begin{array}{c}
f_kf_j\beta_{ki}\beta_{ij}+f_kf_i\beta_{ij}\beta_{jk}+f_if_j\beta_{ik}\beta_
{kj}\\
f_kf_j\partial_i\beta_{kj}+f_kf_i\partial_j\beta_{ik}+f_if_j\partial_k\beta_
{ij}
\end{array}\right.
\label{JH}
\ee
Such relations seem interesting in themselves.
\\[3mm]\indent
Now one can reverse the arguments and, starting with
$G$ satisfying (\ref{WC}), define $\beta_{ij}=
(1/2)[G_{ij}/(G_iG_j)^{1/2}]=\beta_{ji}$.  Then immediately
\be
\partial_k\beta_{ij}=\frac{G_{ijk}}{2(G_iG_j)^{1/2}}-\frac{G_{ij}}
{4(G_iG_j)^{3/2}}\left[G_{ik}G_j+G_iG_{jk}\right]=
\label{KF}
\ee
$$=\frac{1}{4(G_iG_j)^{1/2}}\left[\frac{G_{ik}G_{jk}}{G_k}+\frac{G_{ki}
G_{ij}}{G_i}+\frac{G_{kj}G_{ji}}{G_j}\right]-\frac{G_{ij}}{4(G_iG_j)^{1/2}}
\left[\frac{G_{ik}}{G_i}+\frac{G_{jk}}{G_j}\right]=$$
$$=\frac{G_{ik}G_{jk}}{4G_k(G_iG_j)^{1/2}}=\beta_{ik}\beta_{kj}$$
Also for $f_i=\sqrt{G_i}$ one has
\be
\partial_jf_i=\frac{G_{ij}}{2(G_i)^{1/2}}=\frac{1}{2G_i^{1/2}}2
(G_iG_j)^{1/2}\beta_{ij}=\beta_{ij}f_j
\label{KG}
\ee
which is the reduced DZM system (\ref{WA}).
This shows that (\ref{JB}) characterizes reduced DZM and we have
\\[3mm]\indent {\bf THEOREM 2.1.}$\,\,$
Referring to (\ref{WA}) as the (reduced) DZM system we stipulate
$N$ indices. Then a solution of DZM yields a function $G$ satisfying
(\ref{WC}), (\ref{JF}), (\ref{JG}), and (\ref{JH}) for example.  Conversely
given $G$ satisfying (\ref{WC}) one can define $\beta_{ij}=(1/2)[G_{ij}/
(G_iG_j)^{1/2}]$ such that (\ref{KF}) and 
(\ref{KG}) hold for $f_i=(G_j)^{1/2}$, which
corresponds to reduced DZM. 
\\[3mm]\indent {\bf REMARK 2.2.}$\,\,$
One can also develop an analogy of (\ref{JB}) in a matrix situation
(i.e. $f_j,\,\,\beta_{ij},$ etc. are matrices).  However this requires
commutativity assumptions $f_if_j=f_jf_i$ for example, along with 
$G_{ij}=G_{ji}$ and e.g. $f_k^{-1}G_{ij}=G_{ij}f_k^{-1}$.  We do not
pursue this here.
\\[3mm]\indent
In order to go from (\ref{WC}) to (reduced) DZM to WDVV in a purely algebraic
manner one takes a basis of solutions $f_{jp}$ for $\partial_if_j=
\beta_{ji}f_i$
Here we assume $N$ variables $x^i$ which will be denoted by $u^i$ in
conformity with standard notation involving WDVV and Egorov geometry.
Classical theory cited in \cite{ka} for example yields $N(N-1)/2$ functions
$\beta_{ij}\,\,(i\not=j)$ depending on $N(N-1)/2$ arbitrary functions
of one variable,
along with $N$ functions $f_i$ depending on $N$ arbitrary functions
$\hat{f}_i(u^i)=f_i(0,\cdots,0,u^i,0,\cdots,0)$ determining initial
values.  Then one chooses a basis of solutions $f_{ip},\,\,1\leq p\leq N$,
associated e.g. to $N$ successive choices of $\hat{f}_i$ accompanied
by $N-1$ zeros, and we can write $(\clubsuit)\,\,f_{jp}^2=\partial_jG^p=
G_j^p$.  One should examine this a little more extensively.  Thus
assume the $\beta_{ij}$ are given and consider the equations for $f_i$
in (\ref{WA}) - (\ref{WAA})
(cf. Section 3 and $(\bullet\bullet\bullet)$ therein for more details).
For $N=2$ there is one $\beta_{12}$ and two equations $\partial_1f_2=\beta_
{12}f_1$ and $\partial_2f_1=\beta_{12}f_2$, leading to $(\bullet\bullet)\,\,
\partial_1\partial_2f_2-(\partial_2log\beta_{12})\partial_1f_2-
\beta_{12}^2f_2=0;\,\,
\partial_1\partial_2f_1-(\partial_1log\beta_{12})\partial_2f_1-
\beta_{12}^2f_1=0$
for which a basis of solutions $f_1^1=f_{11},\,\,f_1^2=f_{12},\,\,
f_2^1=f_{21},$ and $f_2^2=f_{22}$ can be envisioned (roughly speaking
these are hyperbolic equations for which two arbitrary functions should 
appear in the integration).  For $N=3$ we have $\beta_{12},\,\,\beta_{13},$
and $\beta_{23}$ with
\be
\partial_1f_2=\beta_{21}f_1;\,\,\partial_1f_3=\beta_{31}f_1;\,\,\partial_2f_3=
\beta_{23}f_2;
\label{WY}
\ee
$$\partial_3f_2=\beta_{23}f_3;\,\,\partial_3f_1=\beta_{13}f_3;\,\,
\partial_2f_1=\beta_{12}f_2$$
leading to equations similar in form to $(\bullet\bullet)$ plus others 
of third order for exmple.  
However it is better
to simply think of (\ref{WY}) as a system of $N=3$ ordinary differential
equations of first order.  To see this we use the stipulation
$\partial f_j=0=(\sum\partial_k)f_j$ from (\ref{WAA}) so that 
for $N=3$ we can write e.g.
\be
\partial_1f_2=\beta_{21}f_1;\,\,\partial_1f_3=\beta_{31}f_1;\,\,
\partial_1f_1=-\partial_2f_1-\partial_3f_1=-\beta_{12}f_2-\beta_{13}f_3
\label{QB}
\ee
or equivalently
\be
\partial_1\left(
\begin{array}{c}
f_1\\
f_2\\
f_3
\end{array}
\right)=\left(
\begin{array}{ccc}
0 & -\beta_{12} & -\beta_{13}\\
\beta_{12} & 0 & 0\\
\beta_{13} & 0 & 0
\end{array}\right)
\left(
\begin{array}{c}
f_1\\
f_2\\
f_3\end{array}
\right)
\label{QC}
\ee
For such systems there is a well known integration theory leading to
a basis $f_{jp}$ as indicated earlier.
\\[3mm]\indent {\bf REMARK 2.3.}$\,\,$
More generally (cf. Section 3) we will want a basis of solutions of
$\partial_if_j=\beta_{ji}f_i$ satisfying $\partial f_j=zf_j$ for a
``spectral parameter" $z$ and the construction is essentially the same.
Thus for $N=3$ for example, in place of (\ref{QB}) - (\ref{QC})
one has $\partial_1f_1=zf_1-\partial_2f_1-\partial_3f_1=zf_1-\beta_{12}f_2
-\beta_{13}f_3$ leading to (\ref{QC}) with a $z$ in the matrix $(1,1)$
position.  In this situation we can provide functions $G^p(z)\,\,(\equiv
G^p(z,t_k)),\,\,1\leq p\leq N$, satisfying (\ref{WC}) with
\be
\partial f_{jp}=(\sum\partial_k)(G^p_j)^{1/2}=\frac{1}{2}\sum_k
(G^p_j)^{-1/2}G^p_{jk}=z(G^p_j)^{1/2}\Rightarrow\sum_k G^p_{jk}=2zG^p_j
\label{QCC}
\ee
For $z=0$ this reduces to $\sum_kG^p_{jk}=0$.
\\[3mm]\indent {\bf REMARK 2.4.}$\,\,$
One (degenerate) road to WDVV goes as follows. Set
\be
c_{ijk}=\sum\left(\frac{G_n^iG_n^jG_n^k}{G^1_n}\right)^{1/2}
\label{ZA}
\ee
The WDVV equations can be written as
\be
\eta^{rs}c_{rjk}c_{\ell ms}=\eta^{rs}c_{rj\ell}c_{kms};\,\,\eta_{ij}=c_{1ij}
=\eta_{ji}=constant
\label{JNN}
\ee 
for which some discussion is given below.  Then putting (\ref{ZA}) in 
(\ref{JNN}) we obtain (cf. $(\clubsuit)$)
\be
\eta^{rs}\sum_n\left(\frac{G^r_nG^j_nG^k_n}{G^1_n}\right)^{1/2}\sum_p
\left(\frac{G^{\ell}_pG^m_pG^s_p}{G^1_p}\right)^{1/2}=
\label{NI}
\ee
$$=\eta^{rs}
\sum_a\left(\frac{G_a^rG_a^jG_a^{\ell}}{G^1_a}\right)^{1/2}\sum_b\left(
\frac{G^k_bG^m_bG^s_b}{G^1_b}\right)^{1/2}$$
Changing indices $a\to n$ and $b\to p$ one has
\be
0=\eta^{rs}\sum\left(\frac{G^r_nG^j_nG^m_pG^s_p}{G^1_nG^1_p}\right)^{1/2}
\left[(G_n^kG_p^{\ell})^{1/2}-(G_n^{\ell}G_p^k)^{1/2}\right]
\label{NJ}
\ee
and one needs conditions which guarantee (\ref{NJ}).  
Assume
there is a functional relation $G^p=G^p(\hat{G})$ for some function
$\hat{G}$ (with no a priori restrictions on $\hat{G}$).  Then
$\partial_n G^k=\partial_{\hat{G}}\partial_n\hat{G}$ and this yields
(we write $G$ for $\hat{G}$ for simplicity)
\be
\frac{G^k_n}{G^k_p}=\frac{\partial_GG^k\partial_nG}{\partial_GG^k
\partial_pG}=\frac{\partial_nG}{\partial_pG}=\frac{\partial_GG^{\ell}
\partial_nG}{\partial_GG^{\ell}\partial_pG}=\frac{G_n^{\ell}}{G^{\ell}_p}
\label{NL}
\ee
which gives (\ref{NJ}).  However we will see in Remark 3.3 that the stipulation
$G^p=G^p(\hat{G})$ leads to a degenerate geometrical situation.

\section{WDVV}
\renewcommand{\theequation}{3.\arabic{equation}}\setcounter{equation}{0}

As background for WDVV we refer
first to \cite{dc} where
one considers a 2D TFT with $N$
primary fields $\phi_i$ and double correlation functions 
$<\phi_i\phi_j>=\eta_{ij}=\eta_{ji}$ (cf. also \cite{fa,ga,kd,mb,mc}
for detailed information on some low dimensional situations).  
The triple correlators $c_{ijk}=
<\phi_i\phi_j\phi_k>$ determine the structure of the operator algebra
of the model via
\be
\phi_i\cdot\phi_j=c_{ij}^k\phi_k;\,\,c^k_{ij}=\eta^{km}c_{ijm};\,\,
(\eta^{ij})=(\eta_{ij})^{-1}
\label{JI}
\ee
along with $<\cdots\phi_i\phi_j\cdots>=c^k_{ij}<\cdots\phi_k\cdots>$.
This is a commutative algebra with unity $\phi_1$ where $c_{1ij}=\eta_{ij}$
with $c^i_{1j}=\delta^i_j$.  The symmetry of $c_{ijk}$ is equivalent to
$<ab,c>=<a,bc>$ and such algebras ($a,b,c\in A$) are called Frobenius
algebras (FA).  It is of concern here to consider algebras $A(t),\,\,
(t=t^1,\cdots,t^N$) with $c_{ijk}=c_{ijk}(t)$ and $\eta_{ij}=constant$
(we assume here that $A(t)$ has no nilpotents - decomposable case).
Then in the TFT situation one can write $c_{ijk}=\partial_i\partial_j
\partial_kF(t)$ for a function $F$ called the primary free energy.
The conditions of associativity give rise to the WDVV equations 
\be
\eta^{\mu\nu}\frac{\partial^3F}{\partial t^{\mu}\partial t^{\beta}
\partial t^{\gamma}}\frac{\partial^3F}{\partial t^{\alpha}\partial t^
{\lambda}\partial t^{\nu}}=\eta^{\mu\nu}\frac{\partial^3F}{\partial t^{\mu}
\partial t^{\beta}\partial t^{\alpha}}\frac{\partial^3F}{\partial t^{\gamma}
\partial t^{\lambda}\partial t^{\nu}}
\label{HH}
\ee
Connections to Egorov geometry arise via the 
Darboux-Egorov (DE) integrable system
(cf. (\ref{WA}))
\be
\partial_k\gamma_{ij}(u)=\gamma_{ik}(u)\gamma_{kj}(u)\,\,(i\not= j
\not= k\,-\,no\,\,sum);\,\,
\partial\gamma_{ij}=(\sum_1^N\partial_k)\gamma_{ij}=0
\label{JK}
\ee
with $\gamma_{ij}(u)=\gamma_{ji}(u)$ for $i\not= j$ (evidently $\gamma_{ij}\sim
\beta_{ij}$ of (\ref{WA}) - (\ref{WAA})). 
The $u^i$ are new
coordinates $u^i=u^i(t)$ defined via
\be
c^k_{ij}(t)=\sum_{m=1}^N\frac{\partial t^k}{\partial u^m}\frac
{\partial u^m}{\partial t^i}\frac{\partial u^m}{\partial t^j}
\label{JL}
\ee
and the functions $\gamma_{ij}(u)$ are expressed via components of the metric
$\eta_{ij}$ where
\be
g_{ij}(u)=\eta_{km}\frac{\partial t^k}{\partial u^i}\frac{\partial t^m}
{\partial u^j}\equiv g_{ii}\delta_{ij};\,\,\gamma_{ij}(u)=\frac
{\partial_j\sqrt{g_{ii}(u)}}{\sqrt{g_{jj}(u)}}=\gamma_{ji}(u)
\label{JM}
\ee
Next one recalls that a diagonal metric $ds^2=\sum_1^Ng_{ii}(u)
(du^i)^2$ determines curvilinear orthogonal coordinates in a Euclidean
space if and only if its curvature vanishes.  This is called an Egorov
metric if the rotation coefficients $\gamma_{ij}=(\partial_j\sqrt{g_{ii}(u)}/
\sqrt{g_{jj}(u)})$ for $(i\not= j)$ satisfy $\gamma_{ij}(u)=\gamma_{ji}(u)$.
Equivalently a potential $V=V(u)$ exists such that $g_{ii}(u)=\partial_i
V(u)\,\,(i=1,\cdots,N)$.  Vanishing of the curvature of the Egorov metric
is equivalent to the integrable system
(\ref{JK}) where (\ref{JK}) corresponds 
to the compatibility
conditions of the system $(\bullet\bullet\bullet)\,\,
\partial_j\psi_i=\gamma_{ij}\psi_j\,\,(i\not= j\,-\,no\,\,sum);
\,\,\partial\psi_j=z\psi_j$,
which is related to the N-wave interaction system.
Note here that compatibility requires
\be
\partial\partial_j\psi_i=(\partial\gamma_{ij})\psi_j+\gamma_{ij}\partial
\psi_j=(\partial\gamma_{ij})\psi_j+z\gamma_{ij}=\partial_j\partial
\psi_i=z\partial_j\psi_i=z\gamma_{ij}\psi_j
\label{JMM}
\ee
so $\partial\gamma_{ij}=0$.
The Egorov zero curvature metric $ds^2$ is called $\partial$-invariant
if $\partial g_{ii}(u)=0$ for $i=1,\cdots,N$ and then it can be specified
uniquely by its rotation coefficients (no arbitrary constants)
via solution of the system
$(\bullet\bullet\bullet)$ for $z=0$ (in particular $\partial\psi_j=0$ as in
(\ref{WAA})).
The same is true for the
corresponding flat coordinates $t^i$.
Thus consider the system $(\spadesuit)\,\,
\partial_j\psi_i=\gamma_{ij}\psi_j$ with $\partial\psi_i=0$
for some solution $\gamma_{ij}=\gamma_{ji}$ of 
$(\bullet\bullet\bullet)$.  Then via
$\gamma_{ij}=\partial_j\sqrt{g_{ii}}/\sqrt{g_{jj}}$ and $\partial g_{ii}=0$
it follows that $\psi_i=\sqrt{g_{ii}}$ is a solution of $(\spadesuit)$.  
Conversely any solution $\psi$ of $(\spadesuit)$ determines a $\partial$
invariant Egorov metric with the same rotation coefficients via 
$g_{ii}=(\psi_{i1})^2$.  Let $\psi_{i1}(u),\cdots,\psi_{iN}$ be a basis
in the space of solutions of $(\spadesuit)$
(cf. Section 2); then one can show that the
scalar product $\eta_{ij}=\sum_1^N\psi_{mi}(u)\psi_{mj}(u)$ is nondegenerate
and does not depend on $u$ (cf. \cite{cz} for details).
The flat coordinates $t^i$ are then determined
by quadratures from the system
\be
\partial_it^j=\psi_{i1}\psi_i^j\equiv\sqrt{g_{ii}}\psi_i^j;\,\,\psi_i^j=
\eta^{jk}\psi_{ik}
\label{JS}
\ee
and one notes that $t_1=\eta_{1i}t^i=V$ is the potential of this metric.
Indeed one obtains
$t^j$ via (\ref{JS}) (i.e. $\partial_it^j=\psi_{i1}\psi^j_i=
\sqrt{g_{ii}}\eta^{jk}\psi_{ik}$) and in particular
\be
\partial_it_1=\partial_i\eta_{1j}t^j=\sqrt{g_{ii}}\eta_{1j}\eta^
{jk}\psi_{ik}=\sqrt{g_{ii}}\psi_{i1}=
g_{ii}=
\psi_{i1}^2=\partial_iV\sim t_1=V
\label{JW}
\ee
Note here from $\partial_j\psi_i=\gamma_{ij}\psi_j$ and $\gamma_{ij}
=\gamma_{ji}$ one has $(\partial_j\psi_i/\psi_j)=(\partial_i\psi_j/\psi_i)$
which implies $\partial_j\psi_i^2=\partial_i\psi_j^2$; hence in particular
this holds for $\psi_j=\psi_{j1}$.
It follows that $ds^2$ is in fact a $\partial$ invariant Egorov metric
of zero curvature.  Finally one proves in \cite{dc} that any solution of
the WDVV equations (in the decomposable case) is determined by a solution
$\gamma_{ij}=\gamma_{ji}$ of the integrable system (\ref{JK}) and by
$N$ arbitrary constants via $g_{ii}=(\psi_{i1})^2,\,\,\eta_{ij}=
\sum_1^N\psi_{mi}\psi_{mj}$, (\ref{JS}), and the formula
\be
c_{ijk}(t)=\sum_1^N\frac{\psi_{mi}\psi_{mj}\psi_{mk}}{\psi_{m1}}
\label{JT}
\ee
One has also the orthogonality
conditions
\be
\frac{\partial t^j}{\partial u^i}=g_{ii}\frac{\partial u^i}{\partial t^k}\eta^
{kj}\sim g_{ii}\frac{\partial u^i}{\partial t^p}=\frac{\partial t^j}{\partial
u^i}\eta_{jp}
\,\,(i=1,\cdots,N)
\label{JU}
\ee
and
the
transformations
\be
\xi_i=\sum_1^Ng^{\frac{1}{2}}_{jj}\frac{\partial u^j}{\partial t^i}\psi_j;\,\,
\psi_i=\sum g_{ii}^{-\frac{1}{2}}\frac{\partial t^j}{\partial u^i}\xi_j
\label{JV}
\ee
Note that since $\psi^2_i=\partial_iV$
one can distinguish between the $\psi_{ip}$ via different
functions $V^p$ as in Theorem 2.1.  
Then for $\psi_{ip}\sim (\partial_iV^p)^{1/2}$ one can define the
$\xi_i^p$ as
\be
\xi_i^p=\sum\psi_{jp}\psi_{ji}=\sum\left(\partial_jV^p
\partial_jV^i\right)^{1/2}
\label{MOO}
\ee
which will agree with the definition via $\xi_i^p=\sum
(\partial u^j/\partial t^i)\psi_{jp}\psi_{j1}$ (cf. (\ref{JU})).
Note also 
\be
\xi_i=g_{ii}^{\frac{1}{2}}\sum g_{jj}^{-\frac{1}{2}}\frac{\partial t^k}{\partial
u^j}\frac{\partial u^j}{\partial t^i}\xi_k
\label{KA}
\ee
\be
u_i^m=\frac{\partial u^m}{\partial t^i}\sim\frac{\psi_{mi}}{\psi_{m1}}
\label{MJ}
\ee
This material is all gathered together in \cite{cz} where many details
are spelled out.
\\[3mm]\indent {\bf REMARK 3.1.}$\,\,$ In connection with the Lam\'e
equations (\ref{WB}) let us note the following.  From $\beta_{ij}=
\partial_jf_i/f_j=(1/2)[G_{ij}/(G_iG_j)^{1/2}]$ we have
\be
\partial\beta_{ij}=\sum_m\beta_{ik}\beta_{kj}=
\partial_i\beta_{ij}+\partial_j\beta_{ji}+\sum_{m\not=i,j}\beta_{mi}
\beta_{mj}
\label{WH}
\ee
$$=
\frac{1}{4(G_iG_j)^{1/2}}\sum_m\left(\frac{G_{mi}G_{mj}}{G_m}\right)=0$$
Consequently
\\[3mm]\indent {\bf THEOREM 3.2.}$\,\,$
The condition $\partial\beta_{ij}=0$ of (\ref{WB}) or (\ref{WH})
corresponds to $B^2=0$ where $B$ is the
matrix $B=(\beta_{ij})$.  Then fitting our functions $G^p$ or $G$ to the
geometric environment involves a stipulation 
\be
0=\sum_m\left(\frac{G_{im}G_{mj}}{G_m}\right)
\label{WI}
\ee
together with (\ref{WC}).
\\[3mm]\indent
{\bf REMARK 3.3.}$\,\,$Let us reconsider the resolution $f^2_{jp}=G^p_j$
with $GP=G^p(\hat{G})$ of Remark 2.3 with regard to the geometrical
meaning connected with $\gamma_{ij}\sim\beta_{ij},\,\,
\psi_j\sim f_j$, and $u^m_j=\partial u^m/\partial t^j\sim (\psi_{mj}/
\psi_{m1})$ from (\ref{MJ}).  One obtains then from (\ref{NL})
\be
\frac{G_n^k}{G^k_p}=\frac{f^2_{nk}}{f^2_{pk}}=\frac{\partial_n\hat{G}}
{\partial_p\hat{G}}=\frac{f^2_{n\ell}}{f^2_{p\ell}}=\frac{G_n^{\ell}}
{G^{\ell}_p}
\label{WJ}
\ee
so (\ref{MJ}) involves
\be
\frac{f_{nk}}{f_{n\ell}}\sim\frac{f_{n1}u^n_k}{f_{n1}u^n_{\ell}}=
\frac{f_{pk}}{f_{p\ell}}=\frac{f_{p1}u^p_k}{f_{p1}u^p_{\ell}}\Rightarrow
\label{WK}
\ee
$$\Rightarrow\frac{u^n_k}{u^n_{\ell}}=\frac{u^p_k}{u^p_{\ell}}\equiv
\frac{u^n_k}{u^p_k}=\frac{u^n_{\ell}}{u^p_{\ell}}=\alpha\equiv
\frac{\partial u^n}{\partial t^k}=\alpha\frac{\partial u^p}{\partial t^k}$$
Hence the Jacobian of transformation $t^i\to u^j$, namely
$J=(\partial(u^j)/\partial(t^i))$ will be ``totally" degenerate.
Consequently $G^p=G^p(\hat{G})$ is not a resolution with significant meaning
and we will now use the underlying geometry to obtain WDVV, while
eventually phrasing matters in terms of $G^p$ (but $G^p\not=
G^p(\hat{G})$).
\\[3mm]\indent
To produce a mechanism leading from DZM to WDVV via the $G^p$ we first
recall a few more formulas from \cite{cz,db,dc}.  First note that 
(\ref{JL}) corresponds to
\be
\frac{\partial u^p}{\partial t^i}\frac{\partial u^p}{\partial t^j}=
c_{ij}^k\frac{\partial u^p}{\partial t^k}
\label{MH}
\ee
since
\be
\sum c^k_{ij}\frac{\partial u^p}{\partial t^k}=\sum\frac{\partial u^m}
{\partial t^i}\frac{\partial u^m}{\partial t^j}\left(\sum\frac{\partial t^k}
{\partial u^m}\frac{\partial u^p}{\partial t^k}\right)
=\frac{\partial u^p}{\partial
t^i}\frac{\partial u^p}{\partial t^j}
\label{JX}
\ee
The WDVV equations (\ref{JNN}) can be also expressed as
\be
c^k_{ij}c^m_{\ell k}=c^k_{\ell j}c^m_{ik}
\label{JPP}
\ee
and this (plus $\partial_ic^m_{jk}=\partial_jc^m_{ik}$) follow as 
compatibility conditions for $(\clubsuit\clubsuit\clubsuit)\,\,
\partial_i\xi_j=z\sum c^k_{ij}\xi_k$.  Here $\xi_j$, as given
in (\ref{JV}) via
$(\spadesuit\spadesuit\spadesuit)\,\,\xi_i=\sum (\partial u^j/\partial t^i)
\psi_j\psi_{j1}$, will not do since (\ref{JV}) corresponds to $z=0$ and we
must go to the $z$ dependent $\psi_j$
(note however that
$z=0$ is needed to produce a uniquely defined zero curvature Egorov
metric).  A direct calculation is conceivable but the argument below
will suffice to
characterize WDVV via the $G^p_j$.
\\[3mm]\indent
First we note that it is really quite wise to follow \cite{db,dc} in
writing $t^{\alpha}$ and $u^i$, with summation on repeated Greek indices
understood.  This will enable us to make sense of formulas involving
derivatives of $\xi_i\sim \xi_{\alpha}$ or $\psi_j$ with respect to
$t^{\beta}$ and $u^i$ together.  In this context then let us isolate some
key features of the WDVV - DZM  correspondence from \cite{db,dc}.  To 
begin we elaborate
on the introduction of the spectral parameter $z$ in $(\bullet\bullet
\bullet)$ and in the equations
\be
\partial_{\alpha}\xi_{\beta}=zc^{\gamma}_{\alpha\beta}(t)\xi_{\gamma}\,\,
(\partial_1\xi_{\beta}=z\xi_{\beta})
\label{QD}
\ee
In an ad hoc spirit an elementary calculation shows that compatibility
of (\ref{QD}) is equivalent to WDVV in the form (\ref{JPP}) along with
$(\bullet\clubsuit\bullet)\,\,\partial_{\delta}c^{\gamma}_{\alpha\beta}=
\partial_{\alpha}c^{\gamma}_{\delta\beta}$.  Similarly compatibility of
$(\bullet\bullet\bullet)$ is equivalent to the DE system (\ref{JK})
(cf. (\ref{JMM})).  Hence let us go to the $\psi_{jp}(z)\sim f_{jp}(z)
\equiv f_{jp}(z,t_k)$ of Remark 2.3 with $G^p_j(z)\equiv G^p_j(z,t_k)\,\,
(1\leq j,p\leq N)$.  From \cite{db,dc} we let
$M$ be a space of parameters
$t^{\alpha}$ and consider a Frobenius algebra (FA) deformation $t^{\alpha}\to
\tilde{t}_{\alpha}(t,z)$ (cf. \cite{db,dc} for the language of FA
and Frobenius manifolds (FM)).  For $\nabla$ the Levi-Civit\`a connection
of the $z=0$ metric one writes $\tilde{\nabla}_u(z)v=\nabla_uv+zu\cdot v
\,\,(\tilde{\nabla}_{\alpha}(z)\sim\partial/\partial\tilde{t}_{\alpha})$
and then $\tilde{\nabla}$ is flat, uniformly in $z$, if and only if
WDVV holds with $c_{\alpha\beta\gamma}=\partial_{\alpha}\partial_{\beta}
\partial_{\gamma}F$.  Here the $c^{\gamma}_
{\alpha\beta}$ are based on the $z=0$ metric and WDVV as in (\ref{JPP})
for example expresses an associativity condition for the corresponding
FA.  Further any $\xi_{\alpha}$ satisfying (\ref{QD}) is a gradient
$\xi_{\alpha}=\partial_{\alpha}\tilde{t}$ for some function $\tilde{t}$ and
one obtains a fundamental system of solutions of (\ref{QD}) via
$\xi_{\alpha}^{\beta}=\partial_{\alpha}\tilde{t}_{\beta}$.  
Actually the $\tilde{t}$ are specified via 
\be
\partial_{\alpha}\partial_{\beta}\tilde{t}=zc^{\epsilon}_{\alpha\beta}
\partial_{\epsilon}\tilde{t}
\label{QE}
\ee
and one will have a coordinate family $\tilde{t}_{\gamma}\,\,(1\leq
\gamma\leq N)$ with $\xi_{\beta}^{\gamma}=\partial_{\beta}\tilde{t}_
{\gamma}$, where $\tilde{t}_{\gamma}=t^{\gamma}+zv^{\gamma}+O(z^2)$.  The
$v^{\gamma}$ are uniquely determined up to a transformation $v^{\gamma}
\to v^{\gamma}+T^{\gamma}_{\beta}t^{\beta}$ for $T$ a constant matrix.
Further there is a formula
\be
z^{\frac{d}{2}-1}\frac{\partial\tilde{t}(t(u),z)}{\psi_{i1}}=\psi_i(u,z)
\label{QF}
\ee
which establishes a $1-1$ correspondence between solutions of 
$(\bullet\bullet\bullet)$ and (\ref{QD}).  Here $d$ is a scaling parameter
arising the consideration of self similar solutions of WDVV and we take it
to be 2 here in order to eliminate the $z$ factor in (\ref{QF}) (cf.
\cite{db,dc,kc,md} for a more comprehensive treatment of WDVV).
\\[3mm]\indent
With this background we pick now $G^p_j(z)\sim f_{jp}^2(z)\sim
\psi_{jp}^2(z)$ as in Remark 2.3, so $(\bullet\bullet\bullet)$ is
satisfied, along with (\ref{JK}).  Note this implies $\partial\beta_{ij}=0$
so (\ref{WH}) is automatic.
From (\ref{QF}) we then obtain $\tilde{t}_{\alpha}(u,z)$
and $\xi_i^{\alpha}$ via
\be
\xi_i^{\alpha}=\partial_i\tilde{t}_{\alpha}=\psi_{i1}\psi_i^{\alpha}=
\psi_{i1}\eta^{\alpha,k}\psi_{ik}
\label{QG}
\ee
which reduces to (\ref{JS}) for $z=0$.  This leads to (\ref{QD}) whose
compatibility implies WDVV.  Hence
\\[3mm]\indent {\bf THEOREM 3.4.}$\,\,$
Let $G^p_j\sim f^2_{jp}\sim\psi^2_{jp}$ be functions of $z$ as in 
Remark 2.3, satisfying (\ref{WC}) and (\ref{QCC}).  Then one obtains
$(\bullet\bullet\bullet)$, and (\ref{JK}) holds for $\gamma_{ij}\sim
\beta_{ij}$, which implies that (\ref{WH}) is automatic.
Further one obtains WDVV with $c_{ijk}$ as in (\ref{ZA}) by means of the
map $\psi_{jk}\to
\xi_i^{\alpha}$ of (\ref{QG}).  This means that such $G^p_j$ characterize
WDVV as well as reduced DZM with (\ref{WAA}).
\\[3mm]\indent {\bf REMARK 3.5.}$\,\,$ For convenience we list together
the stipulations on the $G^p$.  Thus generically 
\be
\begin{array}{cc}
({\bf A}) & \sum_kG_{jk}=2zG_j\\
({\bf B}) & 2G_{ijk}=\frac{G_{ik}G_{kj}}{G_k}+\frac{G_{ki}G_{ij}}{G_i}
+\frac{G_{kj}G_{ji}}{G_j}
\end{array}
\label{QH}
\ee
Then it follows that
$({\bf C}) \,\, \sum_m[(G_{im}G_{mj})/G_m]=0$ and to see this we
note from (\ref{QH}) that
$$
2zG_{ij}=\sum_kG_{ijk}=\frac{1}{2}\left\{\frac{G_{ij}}{G_i}\sum_kG_{ki}
+\frac{G_{ji}}{G_j}\sum_kG_{kj}+\sum_k\frac{G_{ik}G_{kj}}{G_k}\right\}=$$
\be
=2zG_{ij}+\frac{1}{2}\sum_k\frac{G_{ik}G_{kj}}{G_k}
\label{QI}
\ee
In terms of the metric $g_{ii}\sim \psi^2_{i1}\sim f^2_{i1}\sim G^1_i$
the conditions $({\bf A}) - ({\bf B})$ for $G^1$ can be written in the form
\be
2zg_{jj}=\sum_k\partial_kg_{jj};\,\,\,2\partial_i\partial_kg_{jj}=
\frac{\partial_ig_{kk}\partial_jg_{kk}}{g_{kk}}+\frac{\partial_kg_{ii}
\partial_jg_{ii}}{g_{ii}}+\frac{\partial_kg_{jj}\partial_ig_{jj}}{g_{jj}}
\label{QJ}
\ee
where $G^1_{ji}=\partial_ig_{jj}=G^1_{ij}=\partial_jg_{ii}$, etc.
\\[3mm]\indent {\bf ACKNOWLEDGEMENT.}$\,\,$ The author would like to
thank Y. Nutku for stimulating conversations on WDVV and related topics.

\end{document}